\documentclass[a4paper, 11pt]{article}

\pdfoutput=1 

\usepackage{amsmath}
\usepackage[T1]{fontenc} 
\usepackage{authblk}
\usepackage[a4paper, total={6in, 8in}]{geometry}
\usepackage{graphicx}

\begin{document} 

\title{\bf Weak charges quantization in \boldmath $SU(3)_{c} \otimes SU(n)_{L} \otimes U(1)_{Y}$ \bf{gauge models}}

\author{Adrian Palcu}
\affil{"Aurel Vlaicu" University of Arad,\\ 2 Elena Dr\u{a}goi Street, Arad-310330, Romania}
\date{}

\maketitle

\abstract{After proving, in a previous paper, that the electric charge quantization occurs as a natural consequence in renormalizable $SU(3)_c \otimes SU(n)_{L} \otimes U(1)_{Y}$ gauge models, we take here a step further within the same paradigm in order to obtain the precise weak charges quantization. To this end a viable boson mass spectrum yields first, once a proper parametrization in the Higgs sector is employed. Hence, by diagonalizing the neutral bosons mass matrix, the quantized neutral weak charge operators are obtained. The Standard Model phenomenology is entirely recovered, as its scale ($v_{SM}=246$ GeV) is decoupled from the higher scale ($V\sim$ 10 TeV) specific to our generalized electro-weak unification.}

\section{Introduction}
\label{sec:1}
In a recent paper \cite{1} the author worked out in some detail the particular method (conceived by Cot\u{a}escu \cite{2}) for treating the generalized gauge models of the type $SU(3)_{c} \otimes SU(n)_{L} \otimes U(1)_{Y}$, as they could play a decisive role for some of the yet unsolved challenges in particle physics. For some decades now, this sort of models (for the particular cases $n=3,4$) have been regarded in the literature as plausible extensions of the well-established Standard Model (SM) \cite{3}-\cite{5}. The so called 3-3-1 models \cite{6}-\cite{26} or 3-4-1 models \cite{27}-\cite{51} are still extensively investigated in their own right, since some promising results in addressing the tiny neutrino masses, CP issue, dark matter candidates or g-2 muon discrepancy, have already been inferred. Meanwhile, some papers \cite{52}-\cite{54} dared to propose even 3-5-1 models in order to successfully face the available observational data \cite{55}. However, among many shortcomings the theory of the SM contains, one can list, for instance, the lack of predictions regarding the pattern of the electric charge quantization observed in nature and the number of precisely 3 fermion generations. Both these undisputed observational facts can be part of the outcome in each of the particular extensions mentioned above. In Ref.\cite{1} it was elegantly proved, for the most general case, that the renormalization requirement involving the cancellation of the triangle anomalies leads precisely to the true electric charge quantization and to the necessity for 3 fermion  generations (no more, no less). These consequences occur regardless the values $n$ can take, as long as $SU(3)_c$ remains the gauge group of the vector QCD.

What about the couplings for the weak interactions, the so called "weak charges"? Here we deal with this topic, which - we prove straightforwardly in the following - arises as a consequence of the Higgs sector's parametrization, by simply assuming and exploiting the prescriptions of the general method \cite{2}. Obviously, the natural decoupling of the higher scale of the generalized model ($V\sim$ 10 TeV) from the SM's scale ($v_{SM}=246$ GeV) must occur as an undisputed feature of our approach, since the avoidance of any alteration to the SM phenomenology is to be sought. A smart choice of the parameters in the Higgs sector ensures this unequivocally. The resulting values for the weak couplings precisely match the SM ones for the known quarks and leptons, while the predicted values for the exotic quarks and leptons indeed open up a rich phenomenological path to be worked out in future works. 

The paper is organized in 5 sections and a technical appendix. In Section 2 the particle content of the model is briefly reviewed, while the boson mass spectrum is obtained in Section 3, as a consequence of a suitable parametrization in the Higgs sector. Section 4 is properly providing us with the quantized weak charges and is followed by the Section 5 which is reserved for some comments and concluding remarks.

\section{\boldmath $SU(n)_{L} \otimes U(1)_{Y}$ electro-weak sector}
\label{sec:2}

\paragraph{Fermion sector} 
The particle content of the 3-n-1 model under investigation here is displayed below. There are precisely three left-handed generations of leptons and quarks, occurring in the following distinct multiplets (n-plets):

\begin{equation}
L_{iL}=\left(\begin{array}{c}
\vdots \\
N_{i}^{\prime}\\
N_{i}\\
\nu_{i}\\
e_{i}
\end{array}\right)_{L}, \quad Q_{1L}=\left(\begin{array}{c}
\vdots \\
U^{\prime}\\
U\\
u\\
d
\end{array}\right)_{L}  , \quad Q_{2L,3L}=\left(\begin{array}{c}
\vdots \\
D_{2,3}^{\prime}\\
D_{2,3}\\
d_{2,3}\\
u_{2,3}
\end{array}\right)_{L},
\end{equation}
with $i=1,2,3$, and their corresponding right-handed singlet partners.
 
The irreducible representations with respect to the generalized model's gauge group $SU(3)_{c} \otimes SU(n)_{L} \otimes U(1)_{Y}$ stand as:
\begin{equation}
L_{iL}\sim(\boldsymbol{1},\boldsymbol{n},-\frac{1}{n}) \quad , \quad e_{iR}\sim(\boldsymbol{1},\boldsymbol{1},-1) \quad ,\quad \nu_{iR},N_{iR},N_{iR}^{\prime},\ldots \sim(\boldsymbol{1},\boldsymbol{1},0)
\end{equation}

\begin{equation}
Q_{1L}\sim(\boldsymbol{3},\boldsymbol{n},\frac{2n-3}{3n}) \quad , \quad Q_{kL}\sim(\boldsymbol{3},\boldsymbol{n^{*}},\frac{3-n}{3n})
\end{equation}

\begin{equation}
u_{R},u_{kR},U_{R},U_{R}^{\prime},\ldots \sim(\boldsymbol{3},\boldsymbol{1},\frac{2}{3})\quad , \quad d_{R},d_{kR}, D_{kR},D_{kR}^{\prime},\ldots \sim(\boldsymbol{3},\boldsymbol{1},-\frac{1}{3})
\end{equation}
with $k=2,3$.

These assignments are not arbitrary at all, but inferred (see Sec.3.2 in Ref.\cite{1}) by simply requiring to be observed two distinct features: (i) the renormalizability criterion that imposes the cancellation of all the axial anomalies, and (ii) the algebraic constraint on the gauge bosons whose electric charge are allowed to be only $\pm e$ or $0$.   

\paragraph{Gauge sector} 
Now let's turn our attention to the electro-weak interactions in the model at hand. They are mediated by vector bosons defined by the adjoint representation of the (still unbroken) semi-simple gauge group of the electro-weak part of the model.  

That is,
\begin{equation}
A_{\mu}=\left(\begin{array}{ccccccccc}
D_{\mu}^{1}   & \ldots & Y_{\mu}^{\prime\prime0} & Y_{\mu}^{\prime0} & Y_{\mu}^{0} & \vdots \\
\\
\vdots   & \ddots & \vdots & \vdots & \vdots & Y_{\mu}^{\prime\prime+} \\
\\
Y_{\mu}^{\prime\prime0*}  & \ldots  & D_{\mu}^{n-3} & X_{\mu}^{\prime\prime0} & X_{\mu}^{\prime0} & Y_{\mu}^{\prime+}\\
\\
Y_{\mu}^{\prime0*} & \ldots & X_{\mu}^{\prime\prime0*}  & D_{\mu}^{n-2} &  X_{\mu}^{0} & Y_{\mu}^{+} \\
\\
Y_{\mu}^{0*}   & \ldots & X_{\mu}^{\prime0*} & X_{\mu}^{0*}  &  D_{\mu}^{n-1} & W_{\mu}^{+} \\ 
\\
\vdots  &  Y_{\mu}^{\prime\prime-} & Y_{\mu}^{\prime-} & Y_{\mu}^{-} & W_{\mu}^{-}  & D_{\mu}^{n}
\end{array}\right),
\end{equation}
with the diagonal entries (corresponding to the Cartan subalgebra of the $su(n)_{L}\times u(1)_{Y}$ algebra):
 
\begin{equation}
\begin{array}{ll}
D_{\mu}^{1} =&  \frac{1}{2}A_{\mu}^{3}+\frac{1}{2\sqrt{3}}A_{\mu}^{8}+\ldots +YB^{0}_{\mu} \\
\\
D_{\mu}^{2} =  &  -\frac{1}{2}A_{\mu}^{3}+\frac{1}{2\sqrt{3}}A_{\mu}^{8}+\ldots +YB^{0}_{\mu} \\
\\
D_{\mu}^{3} = &  -\frac{1}{\sqrt{3}}A_{\mu}^{8}+\frac{1}{2\sqrt{6}}A_{\mu}^{15}+\ldots +YB^{0}_{\mu} \\
\\
\ldots  \\
\\
D_{\mu}^{n^{2}-1} = & -\frac{\sqrt{n-1}}{\sqrt{2n}}A_{\mu}^{n^{2}-1}+YB^{0}_{\mu}
\end{array}
\end{equation} 

The off-diagonal entries are cast as $B_{\mu}^{\alpha\beta}=\frac{1}{\sqrt{2}}(A_{\mu}^{\alpha}\pm iA_{\mu}^{\beta})$ with $\alpha,\beta=1,2,\ldots, n$, $\alpha\neq\beta$. Evidently, the off-diagonal entries accounts either for charged bosons (if $\alpha=n$ or $\beta=n$), or for neutral bosons (if simultaneously $\alpha\neq n$ and $\beta\neq n$). That means there are gauge bosons exhibiting only $0,\pm$e charges, as we stated above.

\paragraph{Scalar sector}
The SSB is achieved by means of an appropriate scalar sector consisting of the following n scalar multiplets \cite{1}

\begin{equation}
\phi^{(k)}=\left(\begin{array}{l}
\phi_{1}^{(k)}\\
\phi_{2}^{(k)}\\
\vdots\\
\phi_{n}^{(k)}
\end{array}\right)\sim(\boldsymbol{1},\boldsymbol{n},-\frac{1}{n}),  k=1,\ldots,n-1\qquad \phi^{(n)}=\left(\begin{array}{l}
\phi_{1}^{(n)}\\
\phi_{2}^{(n)}\\
\vdots\\
\phi_{n}^{(n)}
\end{array}\right)\sim(\boldsymbol{1},\boldsymbol{n},\frac{n-1}{n})\qquad
\end{equation}
developing each of them its own vacuum expectation value (VEV), in the manner $\left\langle\phi^{(i)}\right\rangle=\eta_{i}$V, due to a set of real parameters $\eta_{i}\in (0,1)$, once a unique overall scale V in the model is assumed. The parameters are cast in a $n \times n$ diagonal matrix ($\eta$) whose trace obey (according to the general method \cite{2}) the constraint $\text{Tr}\left(\eta^{2}\right)$=1. More explicitly,  $\eta^{2}_{1}+\eta^{2}_{2}+\ldots +\eta^{2}_{n}=1$, so that the relation among all $n$ VEVs $\left\langle\phi^{(1)}\right\rangle^{2}+\left\langle\phi^{(2)}
\right\rangle^{2}+\ldots+\left\langle\phi^{(n)}\right\rangle^{2}=V^{2}$ holds. A convenient parameter matrix could be, without loss of generality, the following 
\begin{equation}
\eta^{2}=\text{Diag}\left(\frac{1-a}{n-2},\frac{1-a}{n-2},\ldots \frac{1-a}{n-2},\frac{a-b}{2},\frac{a+b}{2}\right)
\end{equation}  
which, evidently, fulfills the trace requirement. Moreover, it also does split the VEVs, if one considers tuning $a,b \rightarrow 0$ (very small), so that $\left\langle\phi^{(1)}\right\rangle$,$\left\langle\phi^{(2)}\right\rangle$,...,$\left\langle\phi^{(n-2)}\right\rangle\sim V$ and $\left\langle\phi^{(n-1)}\right\rangle$,$\left\langle\phi^{(n)}\right\rangle \sim v_{SM}$. This will play an important role in keeping the SM phenomenology decoupled from the high scale of this model.

\section{Boson mass spectrum}
\label{sec:3}

With the above parameter choice and tuning $a,b\rightarrow0$, one gets (once the SSB is achieved) - according to the prescriptions of the general method \cite{2} - the mass matrix for the Hermitian bosons mediating the weak interactions: 

\begin{equation}
M^{2}=\left(\begin{array}{cccc}
M^{2}(Z_{1}) & 0 & 0 & 0\\
\\
 & \ddots & &
\\
\\
0 & 0 & M^{2}(Z_{n-2}) & 0\\
\\
0 & 0 & 0 & M^{2}_{2\times2}(Z,Z^{\prime})
\end{array}\right)
\end{equation}

As expected, $n-2$ heavier bosons are completely decoupled. Only one, namely ($Z^{\prime}$), mixes with the SM neutral $Z$ boson and could, in principle, interfere with the SM phenomenology. So, only the $2 \times 2$ mass matrix $M^{2}_{2 \times 2}(Z,Z^{\prime})$
\begin{equation}
\frac{2m^{2}}{n-1}\left(\begin{array}{cc}
\frac{1}{n-2}\left[1+\frac{(n^{2}-4n+2)}{2}a-\frac{(n-2)^2}{2}b\right] & \frac{1}{\sqrt{n(n-2)}\cos\theta}\left(1-\frac{n}{2}a+\frac{n-2}{2}b\right)\\
\\
 \frac{1}{\sqrt{n(n-2)}\cos\theta}\left(1-\frac{n}{2}a+\frac{n-2}{2}b\right) & \frac{1}{n\cos^{2}\theta}\left[1+\frac{n(n-2)}{2}a+\frac{n(n-2)}{2}b\right]
\end{array}\right)
\end{equation}
goes actually through the diagonalization procedure. Here $\theta$ represents the rotation angle of a generalized Weinberg transformation (see sec.5 in Ref.\cite{2}) that separates the massless electromagnetic direction from other directions in the parameter space. It is connected to the SM Weinberg angle ($\theta_{W}$) in this way \cite{1}: $\sin\theta=\sqrt{\frac{2(n-1)}{n}}\sin\theta_{W}$.

At this stage one enforces (at the tree level) the mass $m^{2}a/\cos^{2}\theta_{W}$ of the SM neutral boson $Z$  ($\simeq91.2$ GeV \cite{55}) to be an eigenvalue of the matrix in eq.(10). In our parametrization, $M(W^{\pm})=m\sqrt{a}$ ($\simeq80.4$ GeV \cite{55}) by making use of the notation $m^{2}=\frac{1}{4}g^{2}V^{2}$ throughout the proceedings.

Hence, one obtains a nice constraint on the two free parameters $a$ and $b$ (see Appendix below), namely   
\begin{equation}
\left(a\tan^{2}\theta_W+b\right)^2=0
\end{equation}

Under these circumstances the boson mass matrix $M^{2}_{2\times2}(Z,Z^{\prime})$ to be fully diagonalized becomes the one-parameter matrix

\begin{equation}
\frac{2m^{2}}{n-1}\left(\begin{array}{cc}
\frac{1}{n-2}\left[1+a\frac{n^{2}-4n+2+2s^{2}_{W}}{2(1-s^{2}_{W})}\right] & \frac{1}{\sqrt{n(n-2)}\cos\theta}\left[1-a\frac{(n-2s^{2}_{W})}{2(1-s^{2}_{W})}\right]\\
\\
\frac{1}{\sqrt{n(n-2)}\cos\theta}\left[1-a\frac{(n-2s^{2}_{W})}{2(1-s^{2}_{W})}\right]& \frac{1}{n \cos^{2}\theta}\left[1+a\frac{n(n-2)(1-2s^{2})}{2(1-s^{2}_{W})}\right]
\end{array}\right)
\end{equation}

The diagonalization of the matrix in eq.(9) is performed by simply employing the following $SO(n-1)$ matrix, 

\begin{equation}
\omega=\frac{1}{\sqrt{2(n-1)}c_{W}}\left(\begin{array}{ccc}
1_{(n-2)\times(n-2)} & 0 &0 \\ 
0 & -\sqrt{n-2} & \sqrt{n-2(n-1)s^{2}_{W}} \\
0 & -\sqrt{n-2(n-1)s^{2}_{W}} & -\sqrt{n-2} \end{array} \right),
\end{equation}
in the manner
\begin{equation}
\omega M^{2} \omega^{T}=\text{Diag}\left[M^{2}(Z_{1}),M^{2}(Z_{2}),\dots,M^{2}(Z_{n-2}), M^{2}(Z), M^{2}(Z^{\prime})\right].
\end{equation}

The eigenvalues, specific to our 3-n-1 model, are now:
 
\begin{equation}
M^{2}(Z^{\prime})= \frac{4m^{2}(1-s^{2}_{W})}{(n-2)	[n-2(n-1)s^{2}_{W}]}\left[1+a\frac{n(n-4)(1-2s^{2}_{W})-4s^{4}_{W}}{4(1-s^{2}_{W})^{2}}\right],
\end{equation}

\begin{equation}
M^{2}(Z_1)=M^{2}(Z_2)=\ldots=M^{2}(Z_{n-2})=m^{2}\left(\frac{2}{n-2}\right)(1-a),
\end{equation}
all of them much heavier than $Z$, if one considers the parameter $a$ very small, say its order of magnitude $\emph{O}(10^{-3})$ or smaller. Such a tuning is meant to ensure an overall scale $V$ around 10 TeV or higher, according to $v_{SM}=\sqrt{a}V$. 

\section{Weak charges}
\label{sec:4}

Having established the diagonalization matrix $\omega$, all the weak charge operators $Q^{\rho}(Z_{\hat{i}})$ result immediately, according to the prescriptions of the general method Ref.\cite{2}, in this way: 
\begin{equation}
Q^{\rho}(Z_{\hat{i}})=g\left[D_{\hat{k}}^{\rho}-\nu_{\hat{k}}\left(D^{\varrho}\nu\right)(1-\cos\theta)-\nu_{\hat{k}}\frac{g^{\prime}}{g}Y^{\rho}\sin\theta\right]\omega_{\cdot\hat{i}}^{\hat{k}\cdot}
\end{equation}
where $\nu_{\hat{k}}$ are associated to the Hermitian diagonal generators of the gauge group. As it was proved in Ref.\cite{1}, the avoidance of the exotic electric charges implies the selection $\nu_{n^{2}-1}=1$ accompanied simultaneously by the vanishing condition for all the other versors $\nu_{3}=\nu_{8}=\nu_{15}= \dots =0$. That means, in fact, these very versors properly discriminate among the various models based on the same gauge group. 

Thus, the neutral charge operators become, in our particular 3-n-1 model, (up to a $e/s_{w}c_{w}$ factor):
\begin{equation*}
Q^{\rho}(Z)=-\sqrt{\frac{{n-2}}{{2(n-1)}}}T_{n^{2}-2n}^{\rho}+
\end{equation*}

\begin{equation}
+\sqrt{\frac{n-2(n-1)s^{2}_{W}}{2(n-1)}}\left(T_{n^{2}-1}^{\rho}\sqrt{\frac{n-2(n-1)s^{2}_{W}}{n}}-\sqrt{\frac{2(n-1)}{n-2(n-1)s^{2}_{W}}}s^{2}_{W}Y^{\rho}\right),
\end{equation}

\begin{equation*}
Q^{\rho}(Z^{\prime})=-\sqrt{\frac{n-2(n-1)s^{2}_{W}}{2(n-1)}}T_{n^{2}-2n}^{\rho}-
\end{equation*}

\begin{equation}
-\sqrt{\frac{{n-2}}{{2(n-1)}}}\left(T_{n^{2}-1}^{\rho}\sqrt{\frac{n-2(n-1)s^{2}_{W}}{n}}-\sqrt{\frac{2(n-1)}{n-2(n-1)s^{2}_{W}}}s^{2}_{W}Y^{\rho}\right),
\end{equation}

The heavier (decoupled) bosons' couplings yield

\begin{equation}
Q^{\rho}(Z_{1})=gT^{\rho}_{3}=\frac{e}{2s_{W}}\textrm{Diag}(1,-1,0,\ldots,0,0),
\end{equation}

\begin{equation}
Q^{\rho}(Z_{2})=gT^{\rho}_{8}=\frac{e}{2\sqrt{3}s_{W}}\textrm{Diag}(1,1,-2,\ldots,0,0),
\end{equation}

...

\begin{equation}
Q^{\rho}(Z_{n-2})=gT^{\rho}_{n^{2}-4n+3}=\frac{e}{\sqrt{2(n-2)(n-3)}s_{W}}\textrm{Diag}(1,1,\ldots,3-n,0,0),
\end{equation}
where we made use of the notations $s_{W}=\sin\theta_{W}$ and $c_{W}=\cos\theta_{W}$ along with the identification $e=g\sin\theta_{W}$ (once we assumed that the coupling $g$ of the $SU(n)_{L}$ in our model is identical with $g$ of  the $SU(2)_{L}$ in the SM). 

Now, with some simple algebraic computations, the neutral charges are inferred  straightforwardly for all the irreducible representations in our model. For the sake of simplicity, we will express these charges (as usual) in $e/2s_{W}c_{W}$ units, in order to instantly compare them to the well-known SM predicted values. 

The resulting couplings with the SM neutral vector boson $Z$ are

\begin{equation}
Q^{(\boldsymbol{n},-\frac{1}{n})}\left(Z\right)=\left(\begin{array}{ccccc}
0\\
 & \ddots\\
 &  & 0\\
 &  &  & 1\\
 &  &  &  & -1+2s^{2}_{W}
\end{array}\right)
\end{equation}
for the lepton sector, and

\begin{equation}
Q^{(\boldsymbol{n},\frac{2n-3}{3n})}\left(Z\right)=\left(\begin{array}{ccccc}
-\frac{4s^{2}_{W}}{3}\\
 & \ddots\\
 &  & -\frac{4s^{2}_{W}}{3}\\
 &  &  & 1-\frac{4s^{2}_{W}}{3}\\
 &  &  &  & -1+\frac{2s^{2}_{W}}{3}
\end{array}\right),
\end{equation}

\begin{equation}
Q^{(\boldsymbol{n^{*}},\frac{3-n}{3n})}\left(Z\right)=\left(\begin{array}{ccccc}
\frac{2s^{2}_{W}}{3}\\
 & \ddots\\
 &  & \frac{2s^{2}_{W}}{3}\\
 &  &  & -1+\frac{2s^{2}_{W}}{3}\\
 &  &  &  & 1-\frac{4s^{2}_{W}}{3}
\end{array}\right)
\end{equation}
for the quark sector, respectively.

In the case of the new $Z^{\prime}$ neutral vector boson, the couplings are - up to a factor $\frac{\sqrt{n-2}}{\sqrt{n-2(n-1)s^{2}_{W}}}$ - the following:

\begin{equation}
Q^{(\boldsymbol{n},-\frac{1}{n})}\left(Z^{\prime}\right)=\left(\begin{array}{ccccc}
-\frac{2c^{2}_{W}}{n-2}\\
 & \ddots\\
 &  & -\frac{2c^{2}_{W}}{n-2}\\
 &  &  & 1-2s^{2}_{W}\\
 &  &  &  & 1-2s^{2}_{W}
\end{array}\right)
\end{equation}
for the lepton sector, and

\begin{equation}
Q^{(\boldsymbol{n},\frac{2n-3}{3n})}\left(Z^{\prime}\right)=\left(\begin{array}{ccccc}
-\frac{2[3+(1-2n)s^{2}_{W}]}{3(n-2)}\\
 & \ddots\\
 &  & -\frac{2[3+(1-2n)s^{2}_{W}]}{3(n-2)}\\
 &  &  & 1-\frac{2s^{2}_{W}}{3}\\
 &  &  &  & 1-\frac{2s^{2}_{W}}{3}
\end{array}\right),
\end{equation}

\begin{equation}
Q^{(\boldsymbol{n^{*}},\frac{3-n}{3n})}\left(Z^{\prime}\right)=\left(\begin{array}{ccccc}
\frac{2[3-(1+n)s^{2}_{W}]}{3(n-2)}\\
 & \ddots\\
 &  & \frac{2[3-(1+n)s^{2}_{W}]}{3(n-2)}\\
 &  &  & -1+\frac{4s^{2}_{W}}{3}\\
 &  &  &  & -1+\frac{4s^{2}_{W}}{3}
\end{array}\right)
\end{equation}
for the quark sector, respectively.

One can summarize the above obtained results. All the neutral charges of the SM fermions are summarized in Table 1. It is now something of an evidence that the couplings connecting SM-fermions to the neutral SM-vector boson ($Z$) are utterly recovered, meaning that the SM is not altered at all at tree level. At the same time, $Z_{1}$, $\dots$, $Z_{n-2}$ are completely decoupled and exhibit no interactions with the SM fermions. That means those bosons are specific to the our generalized model and have nothing to do with the established SM phenomenology. There is $Z^{\prime}$ that could eventually influence the SM phenomenology. Therefore, our investigation in such models will continue in a future work by estimating the loop corrections that are supposed to provide us with certain restrictions regarding the parameters space.  

\begin{table}[t]
  \begin{center}
    \caption{Couplings of the SM fermions}
    \label{tab:table1}
    \begin{tabular}{l|c|c|c|c|c} 
    \hline      
   couplings $\left(\times\frac{e}{2s_{W}c_{W}}\right)$ & \textbf{$Z$} & \textbf{$Z^{\prime}$} & \textbf{$Z_{1}$} &  $\dots$ & \textbf{$Z_{n-2}$}\\
     \hline
     \hline
     
      $e_{L}$, $\mu_{L}$, $\tau_{L}$ & $-1+2s^{2}_{W}$ & $\frac{(1-2s^{2}_{W})\sqrt{n-2}}{\sqrt{n-2(n-1)s^{2}_{W}}}$ & 0 & $\dots$ & 0 \\
      $\nu_{eL}$, $\nu_{\mu L}$, $\nu_{\tau L}$  & 1 & $\frac{(1-2s^{2}_{W})\sqrt{n-2}}{\sqrt{n-2(n-1)s^{2}_{W}}}$ & 0 & $\dots$ & 0\\
      $e_{R}$, $\mu_{R}$, $\tau_{R}$ & $2s^{2}_{W}$ & $-\frac{2s^{2}_{W}\sqrt{n-2}}{\sqrt{n-2(n-1)s^{2}_{W}}}$ & 0 & $\dots$ & 0 \\
      $\nu_{eR}$, $\nu_{\mu R}$, $\nu_{\tau R}$  & 0 & 0 & 0 & $\dots$ & 0\\
      \hline
      $u_{L}$, $c_{L}$ & $1-\frac{4}{3}s^{2}_{W}$ & $-\frac{(1-\frac{4}{3}s^{2}_{W})\sqrt{n-2}}{\sqrt{n-2(n-1)s^{2}_{W}}}$ & 0 & $\dots$ & 0 \\
      $t_{L}$ & $1-\frac{4}{3}s^{2}_{W}$ & $\frac{(1-\frac{2}{3}s^{2}_{W})\sqrt{n-2}}{\sqrt{n-2(n-1)s^{2}_{W}}}$ & 0 & $\dots$ & 0 \\
      $d_{L}$, $s_{L}$  & $-1+\frac{2}{3}s^{2}_{W}$ & $-\frac{(1-\frac{4}{3}s^{2}_{W})\sqrt{n-2}}{\sqrt{n-2(n-1)s^{2}_{W}}}$ & 0 & $\dots$ & 0\\
      $b_{L}$  & $-1+\frac{2}{3}s^{2}_{W}$ & $\frac{(1-\frac{2}{3}s^{2}_{W})\sqrt{n-2}}{\sqrt{n-2(n-1)s^{2}_{W}}}$ & 0 & $\dots$ & 0\\
      $u_{R}$, $c_{R}$, $t_{R}$ & $-\frac{4}{3}s^{2}_{W}$ & $\frac{4s^{2}_{W}\sqrt{n-2}}{3\sqrt{n-2(n-1)s^{2}_{W}}}$ & 0 & $\dots$ & 0 \\
      $d_{R}$, $s_{R}$, $b_{R}$  & $\frac{2}{3}s^{2}_{W}$ & $-\frac{2s^{2}_{W}\sqrt{n-2}}{3\sqrt{n-2(n-1)s^{2}_{W}}}$ & 0 & $\dots$ & 0\\
      \hline
      \hline
    \end{tabular}
  \end{center}
\end{table}

For the heavier $Z_{1}$, $Z_{2}$, $\ldots$, $Z_{n-2}$ neutral bosons the computation is actually simpler. Their couplings are given by their associated diagonal generators in the $su(n)\otimes u(1)$ algebra and nothing else. That is, only $T_{3}$ accounts for $Z_{1}$, only $T_{8}$ accounts for $Z_{2}$ couplings, and so forth ..., with no admixture at all. The resulting values are summarized in Table 2, once the explicit expressions are given below.

The couplings for $Z_{1}$ yield:

\begin{equation}
Q^{(\boldsymbol{n},-\frac{1}{n})}\left( Z_{1}\right)=c_{W}\left(\begin{array}{ccccc}
1\\
 & -1\\
 &  & 0\\
 &  &  & \ddots\\
 &  &  &  & 0
\end{array}\right)
\end{equation}
for the lepton sector, and

\begin{equation}
Q^{(\boldsymbol{n},\frac{2n-3}{3n})}\left( Z_{1}\right)=c_{W}\left(\begin{array}{ccccc}
1\\
 & -1\\
 &  & 0\\
 &  &  & \ddots\\
 &  &  &  & 0
\end{array}\right)
\end{equation}

\begin{equation}
Q^{(\boldsymbol{n^{*}},-\frac{n-3}{3n})}\left( Z_{1}\right)=c_{W}\left(\begin{array}{ccccc}
-1\\
 & 1\\
 &  & 0\\
 &  &  & \ddots\\
 &  &  &  & 0
\end{array}\right)
\end{equation}
for the quark sector, respectively. 

In the case of the $Z_{2}$, the couplings are inferred as:

\begin{equation}
Q^{(\boldsymbol{n},-\frac{1}{n})}\left( Z_{2}\right)=\frac{c_{W}}{\sqrt{3}}\left(\begin{array}{cccccc}
1\\
 & 1\\
 &  & -2\\
 &  &  & 0\\
 &  &  &  & \ddots\\
 &  &  &  &  &  0\\
\end{array}\right)
\end{equation}
for the lepton sector, and

\begin{equation}
Q^{(\boldsymbol{n},\frac{2n-3}{3n})}\left( Z_{2}\right)=\frac{c_{W}}{\sqrt{3}}\left(\begin{array}{cccccc}
1\\
 & 1\\
 &  & -2\\
 &  &  & 0\\
 &  &  &  & \ddots\\
 &  &  &  &  &  0\\
\end{array}\right)
\end{equation}

\begin{equation}
Q^{(\boldsymbol{n^{*}},\frac{3-n}{3n})}\left( Z_{2}\right)=\frac{c_{W}}{\sqrt{3}}\left(\begin{array}{cccccc}
-1\\
 & -1\\
 &  & 2\\
 &  &  & 0\\
 &  &  &  & \ddots\\
 &  &  &  &  &  0\\
\end{array}\right)
\end{equation}
for the quark sector, respectively.

The couplings for $Z_{n-2}$ yield:

\begin{equation}
Q^{(\boldsymbol{n},-\frac{1}{n})}\left( Z_{n-2}\right)=\frac{\sqrt{2}c_{W}}{\sqrt{(n-2)(n-3)}}\left(\begin{array}{ccccccc}
 1\\
 &  1\\
 &  &  1\\
 &  &  & \ddots\\
 &  &  &  & 3-n\\
 &  &  &  &  &  0\\
 &  &  &  &  &  &  0\\
\end{array}\right)
\end{equation}
for the lepton sector, and

\begin{equation}
Q^{(\boldsymbol{n},\frac{2n-3}{3n})}\left( Z_{n-2}\right)=\frac{\sqrt{2}c_{W}}{\sqrt{(n-2)(n-3)}}\left(\begin{array}{ccccccc}
 1\\
 &  1\\
 &  &  1\\
 &  &  & \ddots\\
 &  &  &  & 3-n\\
 &  &  &  &  &  0\\
 &  &  &  &  &  &  0\\
\end{array}\right)
\end{equation}

\begin{equation}
Q^{(\boldsymbol{n^{*}},-\frac{n-3}{3n})}\left( Z_{n-2}\right)=\frac{\sqrt{2}c_{W}}{\sqrt{(n-2)(n-3)}}\left(\begin{array}{ccccccc}
 -1\\
 &  -1\\
 &  &  -1\\
 &  &  & \ddots\\
 &  &  &  & n-3\\
 &  &  &  &  &  0\\
 &  &  &  &  &  &  0\\
\end{array}\right)
\end{equation}
for the quark sector, respectively.

Let's turn now to the fermion singlets. Their weak charges are computed in the simplest way, since one takes into account only the $U(1)$ hypercharges $Y$ and none of the $T$ generators. Hence, for the right-handed singlets, there are no interactions at all  with the heavier $Z_{1}$, $Z_{2}$, ..., $Z_{n-2}$. At the same time, as expected, all the neutral right-handed fermions have no weak interactions regardless the SM or non-SM bosons.  

\begin{table}
  \begin{center}
    \caption{Couplings of the non-SM fermions}
    \label{tab:table2}
    \begin{tabular}{l|c|c|c|c|c|c}
    \hline       
   couplings $\left(\times\frac{e}{2s_{W}c_{W}}\right)$ & \textbf{$Z$} & \textbf{$Z^{\prime}$} & \textbf{$Z_{1}$} & $\dots$ & \textbf{$Z_{n-3}$} & \textbf{$Z_{n-2}$}\\
     \hline
     \hline
     
      $N_{eL}$, $N_{\mu L}$, $N_{\tau L}$ & 0 & $\frac{-2c^{2}_{W}}{\sqrt{(n-2)[n-2(n-1)s^{2}_{W}}]}$ & 0 & $$\dots$$ & 0 & $\frac{-c_{W}\sqrt{2(n-3)}}{\sqrt{n-2}}$\\
      $N^{\prime}_{eL}$, $N^{\prime}_{\mu L}$, $N^{\prime}_{\tau L}$  & 0 & $\frac{-2c^{2}_{W}}{\sqrt{(n-2)[n-2(n-1)s^{2}_{W}}]}$ & 0 & $\dots$ & $\frac{-c_{W}\sqrt{2(n-4)}}{\sqrt{n-3}}$ & $\frac{c_{W}\sqrt{2}}{\sqrt{(n-2)(n-3)}}$\\
      $N^{\prime\prime}_{eL}$, $N^{\prime\prime}_{\mu L}$,   $N^{\prime\prime}_{\tau L}$  & 0 & $\frac{-2c^{2}_{W}}{\sqrt{(n-2)[n-2(n-1)s^{2}_{W}}]}$ & 0 & $\dots$ & $\frac{c_{W}\sqrt{2}}{\sqrt{(n-3)(n-4)}}$ & $\frac{c_{W}\sqrt{2}}{\sqrt{(n-2)(n-3)}}$\\
      $\dots$ & & & & & & \\
      $N_{eR}$, $N_{\mu R}$, $N_{\tau R}$ & 0 & 0 & 0 & $\dots$ & 0 & 0 \\
      $N^{\prime}_{eR}$, $N^{\prime}_{\mu R}$, $N^{\prime}_{\tau R}$  & 0 & 0 & 0 & $\dots$ & 0 & 0\\
      $N^{\prime\prime}_{eR}$, $N^{\prime\prime}_{\mu R}$, $N^{\prime\prime}_{\tau R}$  & 0 & 0 & 0 & $\dots$ & 0 & 0\\
      $\dots$ & & & & & & \\
      \hline         
      $U_{1L}$  & $-\frac{4}{3}s^{2}_{W}$ & $-\frac{2[3+(1-2n)s^{2}_{W}]}{3\sqrt{(n-2)[n-2(n-1)s^{2}_{W}]}}$ & 0 & $\dots$ & 0 & $\frac{-c_{W}\sqrt{2(n-3)}}{\sqrt{n-2}}$ \\
      
      $D_{2L}$, $D_{3L}$ & $\frac{2}{3}s^{2}_{W}$ & $\frac{2[3-(1+n)s^{2}_{W}]}{3\sqrt{(n-2)[n-2(n-1)s^{2}_{W}]}}$ & 0 & $\dots$ & 0 & $\frac{c_{W}\sqrt{2(n-3)}}{\sqrt{n-2}}$ \\
      $U^{\prime}_{1L}$  & $-\frac{4}{3}s^{2}_{W}$ & $-\frac{2[3+(1-2n)s^{2}_{W}]}{3\sqrt{(n-2)[n-2(n-1)s^{2}_{W}]}}$ & 0 & $\dots$ &  $\frac{-c_{W}\sqrt{2(n-4)}}{\sqrt{n-3}}$ & $\frac{c_{W}\sqrt{2}}{\sqrt{(n-2)(n-3)}}$ \\
     
      $D^{\prime}_{2L}$, $D^{\prime}_{3L}$  & $\frac{2}{3}s^{2}_{W}$ & $\frac{2[3-(1+n)s^{2}_{W}]}{3\sqrt{(n-2)[n-2(n-1)s^{2}_{W}]}}$  & 0 & $\dots$ & $\frac{c_{W}\sqrt{2(n-4)}}{\sqrt{n-3}}$ &$\frac{-c_{W}\sqrt{2}}{\sqrt{(n-2)(n-3)}}$ \\
      
      $U^{\prime\prime}_{1L}$ & $-\frac{4}{3}s^{2}_{W}$ & $-\frac{2[3+(1-2n)s^{2}_{W}]}{3\sqrt{(n-2)[n-2(n-1)s^{2}_{W}]}}$ &  0  & $\dots$  & $\frac{c_{W}\sqrt{2}}{\sqrt{(n-3)(n-4)}}$ & $\frac{c_{W}\sqrt{2}}{\sqrt{(n-2)(n-3)}}$ \\

      $D^{\prime\prime}_{2L}$, $D^{\prime\prime}_{3L}$ & $\frac{2}{3}s^{2}_{W}$ & $\frac{2[3-(1+n)s^{2}_{W}]}{3\sqrt{(n-2)[n-2(n-1)s^{2}_{W}]}}$  & 0 & $\dots$ & $\frac{-c_{W}\sqrt{2}}{\sqrt{(n-3)(n-4)}}$ & $\frac{-c_{W}\sqrt{2}}{\sqrt{(n-2)(n-3)}}$ \\
      $\dots$ & & & & & & \\
      $U_{1R}$, $U^{\prime}_{1R}$, $U^{\prime\prime}_{1R}$, $\dots$ & $-\frac{4}{3}s^{2}_{W}$ & $\frac{4s^{2}_{W}\sqrt{n-2}}{3\sqrt{n-2(n-1)s^{2}_{W}}}$ & 0 & $\dots$ & 0 & 0 \\
      $D_{kR}$, $D^{\prime}_{kR}$, $D^{\prime\prime}_{kR}$, $\dots$ & $\frac{2}{3}s^{2}_{W}$ & $-\frac{2s^{2}_{W}\sqrt{n-2}}{3\sqrt{n-2(n-1)s^{2}_{W}}}$ & 0 & $\dots$ & 0 & 0 \\
    
        \hline
        \hline
    \end{tabular}
  \end{center}
\end{table}

For the sake of completeness we display below the weak interactions of the right handed representations explicitly. In the case of charged leptons one obtains the following couplings:
\begin{equation}
Q^{(\boldsymbol{1},-1)}\left( Z\right)=2s^{2}_{W}, \quad
Q^{(\boldsymbol{1},-1)}\left( Z^{\prime}\right)=-\frac{2\sqrt{n-2}s^{2}_{W}}{\sqrt{n-2(n-1)s^{2}_{W}}},
\end{equation}
while all kinds of right-handed neutrinos are sterile  
\begin{equation}
Q^{(\boldsymbol{1},0)}\left( Z\right)=0, \quad
Q^{(\boldsymbol{1},0)}\left( Z^{\prime}\right)=0,
\end{equation}
as expected.

In the quark sector, the right-handed up-type quarks couplings yield
\begin{equation}
Q^{(\boldsymbol{1},2/3)}\left( Z\right)=-\frac{4s^{2}_{W}}{3},\quad
Q^{(\boldsymbol{1},2/3)}\left( Z^{\prime}\right)=\frac{4\sqrt{n-2}s^{2}_{W}}{3\sqrt{n-2(n-1)s^{2}_{W}}},
\end{equation}
while the right-handed down-type quarks interacts weakly in the manner
\begin{equation}
Q^{(\boldsymbol{1},-1/3)}\left( Z\right)=\frac{2s^{2}_{W}}{3},\quad
Q^{(\boldsymbol{1},-1/3)}\left( Z^{\prime}\right)=-\frac{2\sqrt{n-2}s^{2}_{W}}{3\sqrt{n-2(n-1)s^{2}_{W}}}.
\end{equation}

\section{Concluding remarks} 
\label{sec:5}
The starting point of our discussion here was the quantized electric charge operator in $SU(3)_{c}\otimes SU(n)_{L}\otimes U(1)_{Y}$ generalized gauge models, proved in Ref.\cite{1} to match the exact electric charge pattern observed in nature. This was achieved based only on renormalization criteria imposed to the chiral anomalies, with no supplemental hypothesis or computational artifices. Here we enlarged our approach and inferred the quantization of all the weak charge operators in model, as a mere consequence of the same renormalizable paradigm. Moreover, by combining eqs.(18)-(19) with the generalized Gell-Mann--Nishijima formula (eq.(13) in Ref.\cite{1}) in order to get rid of the would-be hypercharges $Y$, one obtains (up to a factor $\frac{e}{s_{W}c_{W}}$) the following expressions:
\begin{equation}
Q^{\rho}(Z)=-T_{n^{2}-2n}^{\rho}\sqrt{\frac{{n-2}}{{2(n-1)}}}+T_{n^{2}-1}^{\rho}\frac{\sqrt{n}}{\sqrt{2(n-1)}}-s^2_{W}Q^{\rho}_{em},
\end{equation}

\begin{equation*}
Q^{\rho}(Z^{\prime})=-T_{n^{2}-2n}^{\rho}\sqrt{\frac{n-2(n-1)s^{2}_{W}}{2(n-1)}}-T_{n^{2}-1}^{\rho}\frac{\sqrt{n(n-2)}}{\sqrt{2(n-1)[n-2(n-1)s^{2}_{W}]}}
\end{equation*}

\begin{equation}
+Q^{\rho}_{em}s^2_{W}\sqrt{\frac{{n-2}}{n-2(n-1)s^{2}_{W}}},
\end{equation}
for $Z$ and $Z^{\prime}$ couplings, while the rest of the $n-2$ weak interactions can be picked up from eqs.(20)-(22). It goes without saying that the weak charges are quantized, once the electric charge is quantized and they are strictly related for each irreducible representation $\rho$, in the manner displayed above.  

This is a remarkable outcome even though, for the moment, our discussion regards the tree level only. Any quantum correction remains to be worked out, once a specific process is taken into consideration. As one can easily observe from the results derived in the present work, the SM-fermions preserve their couplings predicted by the SM. Moreover, they have no couplings at all with the new bosons $Z_{1}$, $Z_{2}$, ..., $Z_{n-2}$, but only with $Z^{\prime}$ ($\geq5.1$ TeV \cite{55}) whose influence must be estimated in a future work. If worked out properly such corrections could enforce restrictions on the overall scale of the model and other parameters as well (for instance, the appropriate Yukawa couplings in the neutrino sector or quark sector). At the same time, a nice feature of our generalized gauge model resides in the fact that the SM-boson $Z$ has vector interactions with all heavier fermions (other than SM-fermions), while $Z^{\prime}$ makes no distinction on the electric charge basis when it comes to the left-handed fermions in the same SM doublet. A plethora of suitable candidates for the cold Dark Matter can be found in such models, since a lot of neutral fermions, neutral scalars and even neutral vector bosons are not interacting with ordinary matter particles - as the new scale $V$ is actually decoupled from the SM's scale. Furthermore, when one goes to particular realistic models such as 3-3-1, 3-4-1 or 3-5-1 (the only three kinds of possible SM extensions compatible with the QCD based on $SU(3)$ - see discussion in Ref.\cite{1}), one recovers the proper charge spectrum by simply applying the prescriptions obtained in the general method. Last but not least, we could advocate an additional argument in favor of firmly considering the generalized $SU(3)_{c} \otimes SU(n)_{L} \otimes U(1)_{Y}$ gauge models. For the time being, this could seem only a hypothetical speculation, but the history of science proved many times that it is not wise to discard even the weirdest hypothesis. What if the QCD one day will be reassessed on a larger gauge group than $SU(3)_{c}$? Then the number of colors could be more than three, and the asymptotic freedom could restrict the number of quark flavors to a much larger number than 16, as it results now from the renormalization group procedure. Consequently, the electro-weak sector could be based on larger $SU(n)_{L} \otimes U(1)_{Y}$ group and our approach becomes a very useful tool.      

\appendix
\section{Diagonalization of the boson mass matrix}

In order to get $M^{2}(Z)=m^{2}a/c^{2}_{W}$ as an eigenvalue of the matrix in eq.(10), one has to compute the characteristic equation, namely the determinant

\begin{equation}
\left|\begin{array}{cc}
\frac{1}{n-2}\left[1+\frac{(n^{2}-4n+2)}{2}a-\frac{(n-2)^2}{2}b\right]-\frac{a(n-1)}{2(1-s^{2}_{W})} & \frac{1}{\sqrt{n(n-2)}\cos\theta}\left(1-\frac{n}{2}a+\frac{n-2}{2}b\right)\\
\\
 \frac{1}{\sqrt{n(n-2)}\cos\theta}\left(1-\frac{n}{2}a+\frac{n-2}{2}b\right) & \frac{1}{n\cos^{2}\theta}\left[1+\frac{n(n-2)}{2}a+\frac{n(n-2)}{2}b\right]-\frac{a(n-1)}{2(1-s^{2}_{W})}
\end{array}\right|
\end{equation}
which has to be equal to zero. Thus, one will be led to a particular relation between the two free parameters. 

The computation reduces to simply summing the following coefficients of the powers of the parameters. They are in order

free term: 

\begin{equation}
\frac{1}{n(n-2)\cos^{2}\theta} - \frac{1}{n(n-2)\cos^{2}\theta}=0
\end{equation}

$a$: 
\begin{equation}
\frac{-n-(n^{2}-4n+2)s^{2}_{W}-n+(n^{2}-2n+2)s^{2}_{W}+2n(1-s^{2}_{W})}{2n(n-2)\cos^{2}\theta(1-s^{2}_W)}=0
\end{equation}

$b$: 
\begin{equation}
\frac{-(n-2)^{2}+n(n-2)-2(n-2)}{2n(n-2)\cos^{2}\theta}=0
\end{equation}

$ab$: 
\begin{equation*}
-\frac{(n-2)\lbrace n[n+(n^{2}-4n+2)s^{2}_{W}] - (n-2)[n-(n^{2}-2n+2)s^{2}_{W}]-2n(1-s^{2}_{W})\rbrace}{4n(n-2)\cos^{2}\theta(1-s^{2}_{W})}
\end{equation*}

\begin{equation}
\ldots=-\frac{(n-2)(n-1)^{2}}{4n\cos^{2}\theta}\frac{2s^{2}_{W}}{(1-s^{2}_{W})}
\end{equation}

$a^2$: 
\begin{equation*}
\frac{\lbrace [n+(n^{2}-4n+2)s^{2}_{W}] [n-(n^{2}-2n+2)s^{2}_{W}]-n^{2}(1-s^{2}_{W})^{2}\rbrace}{4n(n-2)(1-s^{2}_{W})^{2}\cos^{2}\theta}
\end{equation*}

\begin{equation}
\ldots=-\frac{(n-2)(n-1)^{2}}{4n\cos^{2}\theta}\frac{s^{4}_{W}}{(1-s^{2}_{W})^{2}}
\end{equation}

$b^2$: 
\begin{equation}
-\frac{[(n-2)^{3} n + (n-2)^{2} ]}{4n(n-2)\cos^{2}\theta}=-\frac{(n-2)(n-1)^{2}}{4n\cos^{2}\theta}
\end{equation}

Now, the characteristic equation becomes 

\begin{equation}
\left(a\frac{s^{2}_{W}}{1-s^{2}_{W}}+b\right)^2=0
\end{equation}
and it actually removes one parameter (say $b$), so that finally one remains with a mass matrix depending only on the parameter $a$.

\end{document}